\documentclass[aip,apl,epsfig,showpacs,twocolumn]{revtex4}

\usepackage{graphicx}

\begin{document}

\title{Nearly total spin polarization in La$_{2/3}$Sr$_{1/3}$MnO$_3$ from tunneling experiments}

\author{M. Bowen}
\author{M. Bibes}
\author{A. Barth\'el\'emy}
\author{J.-P. Contour}
\author{A. Anane}
\author{Y. Lema\^{\i}tre}
\author{A. Fert}

\affiliation{Unit\'e Mixte de Physique CNRS / Thales, Domaine de
Corbeville, 91404 Orsay, France and Universit\'e de Paris-Sud,
91405 Orsay, France}

\date{\today}

\vspace{0.5cm}

\begin{abstract}

\vspace{0.5cm}

We have performed magnetotransport measurements on
La$_{2/3}$Sr$_{1/3}$MnO$_3$ / SrTiO$_3$ /
La$_{2/3}$Sr$_{1/3}$MnO$_3$ magnetic tunnel junctions. A
magnetoresistance ratio of more than \mbox{1800 \%} is obtained
at 4K, from which we infer an electrode spin polarization of at
least \mbox{95 \%}.  This result strongly underscores the
half-metallic nature of mixed-valence manganites and demonstrates
its capability as a spin analyzer. The magnetoresistance extends
up to temperatures of more than 270K. We argue that these
improvements over most previous works may result from optimizing
the patterning process for oxide heterostructures.

\end{abstract}

\maketitle

Magnetic tunnel junctions (MTJ) have been studied actively from
the mid 90's \cite{moodera95} due to both the underlying physics
and their potential applications as magnetic memories (MRAMs) or
sensors. These structures consist of two ferromagnetic metallic
electrodes (FM) sandwiching a thin insulating barrier (I). When a
bias voltage V$\rm{_{DC}}$ is applied, electrons near the FM/I
interface tunnel through the barrier and, since they are
spin-polarized, the resistance depends on the relative
orientation of the electrodes' magnetization. The tunneling
magnetoresistance (TMR) ratio is defined as

\begin{equation}
\rm{TMR = (R_{AP}-R_P)/R_P } \label{tmr}
\end{equation}

\noindent where R$\rm{_{AP}}$ and R$\rm{_{P}}$ are the resistances
of the junction in the antiparallel and parallel configurations,
respectively. In Julliere's model \cite{julliere75}, the TMR ratio
is related to the spin polarizations P$_1$ and P$_2$ of the two
ferromagnetic electrodes as:

\begin{equation}
\rm{TMR = 2P_1 P_2 /(1-P_1 P_2) } \label{julliere}
\end{equation}

Within this simple model, large TMR ratios result from electrodes,
or from electrode-barrier interfaces \cite{deteresa99b}, with
large effective spin polarization values. Junctions which
integrate amorphous barriers such as Al$_2$O$_3$ and transition
ferromagnets, for which the spin polarization does not exceed
around 50 \% \cite{moodera95,monsma2000}, do not show TMR ratios
larger than 60 \%. Preliminary work has been
reported\cite{bowen2001} on obtaining large interfacial spin
polarizations owing to band structure effects, but the simplest
route to achieving large TMR ratios relies on the use of
so-called "half-metals" with a nearly total intrinsic spin
polarization.

A few compounds have been predicted to be half-metallic, such as
CrO$_2$ \cite{lewis97}, Fe$_3$O$_4$ \cite{zhang91}, mixed-valence
manganites \cite{pickett96} or some Heusler alloys
\cite{degroot83}. In the particular case of manganites such as
La$_{2/3}$Sr$_{1/3}$MnO$_3$ (LSMO) and
La$_{2/3}$Ca$_{1/3}$MnO$_3$ (LCMO), there is a lot of controversy
regarding their half-metallicity. Indeed, whereas spin-polarized
photoemission spectroscopy experiments \cite{park98b} have
confirmed the half-metallic character of LSMO, the maximum spin
polarization as inferred from tunneling experiments does not
exceed 86\% in LCMO \cite{jo2000} and 83\% in LSMO \cite{viret97}.

In this letter, we report a TMR ratio of more than \mbox{1800 \%}
at T=4.2K and V$\rm{_{DC}}$=1mV in La$_{2/3}$Sr$_{1/3}$MnO$_3$ /
SrTiO$_3$ / La$_{2/3}$Sr$_{1/3}$MnO$_3$ fully epitaxial MTJs,
from which we deduce a spin polarization of at least 95 \% for
LSMO. This result confirms for the first time the transport
half-metallic nature \cite{mazin99} of this material, which can
therefore be used as a spin analyzer in tunneling
experiments\cite{deteresa99b}. We argue that this large TMR value
arises both from preserving the quality of the LSMO / STO (STO :
SrTiO$_3$) interfaces during our upgraded patterning process, and
from designing junctions of small size. The TMR extends to
temperatures of about 280K, an improvement compared to previous
results in the literature \cite{obata99}.

\begin{figure}
 \includegraphics[width=0.9\columnwidth]{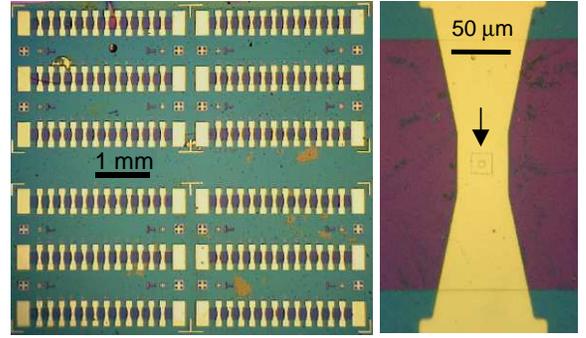}
 \caption{Optical images of the processed sample}
 \label{litho}
\end{figure}

LSMO 350 $\rm{\AA}$/ STO 28 $\rm{\AA}$/ LSMO 100 $\rm{\AA}$
epitaxial trilayer structures were grown by pulsed laser
deposition onto (001)-oriented 10 mm $\times$ 10 mm $\times$ 0.5
mm STO commercial substrates in the same conditions as in
references \onlinecite{viret97} and \onlinecite{lyonnet2000}.
With an idea to induce a pinning effect on the top LSMO
electrode, samples were inserted into a r.f. sputtering system
and 150 $\rm{\AA}$ of Co were deposited, then etched by an
oxygen-rich plasma to form a CoO layer some 25 $\rm{\AA}$ thick.
The samples were finally capped with 150 $\rm{\AA}$ of gold. The
LSMO in-plane cell parameters as measured by X-ray diffraction
are equal to those of STO, implying that the oxide part of the
heterostructure is unrelaxed from the elastic point of
view\cite{lyonnet2000}. Scanning electron energy loss spectroscopy
(EELS) experiments performed at the atomic scale on samples grown
in the same conditions revealed only a very weak modification of
the electronic properties of the LSMO at the interface with the
STO barrier when compared to regions located deeper inside the
LSMO layer \cite{pailloux2001} (no change in the valence of the Mn
ions).

The patterning process was carried out by standard UV
photolithography techniques \cite{montaigne98} using chromium
masks which define 144 MTJs ranging in size from 2 to 6144
$\mu$m$^2$, within a 6 mm $\times$ 6 mm surface. In the first
step, 144 pillars were defined by photolithography and ion-beam
etching. During the etching process, the sample was mounted on a
water-cooled sample holder. Ar ions were accelerated with a grid
voltage of 200 V and neutralized by an electron-emitting
filament. The etching process was monitored with an in-situ
Secondary Ion Mass Spectroscope so as to stop the etching when
entering the bottom LSMO layer. In the second step, twelve 200
$\mu$m-wide bottom electrodes were created using the same
combination of photolithography and neutralized ion-beam etching.
To passivate the sample, a 2500 $\rm{\AA}$-thick layer of
Si$_3$N$_4$ was deposited by dc-sputtering and selectively
removed by reactive ion etching to define electrical access
points. Finally, Ti/Au tracks were deposited as electrical
contacts for transport measurements. Optical images of the
completed sample are shown in Figure \ref{litho}.  After
completing the patterning process, MTJ resistance was checked at
room temperature, which revealed that the junctions of area larger
than 32 $\mu$m$^2$ were short-circuited. Among the remaining
ones, 12 could be measured.

\begin{figure}
 \includegraphics[width=0.9\columnwidth]{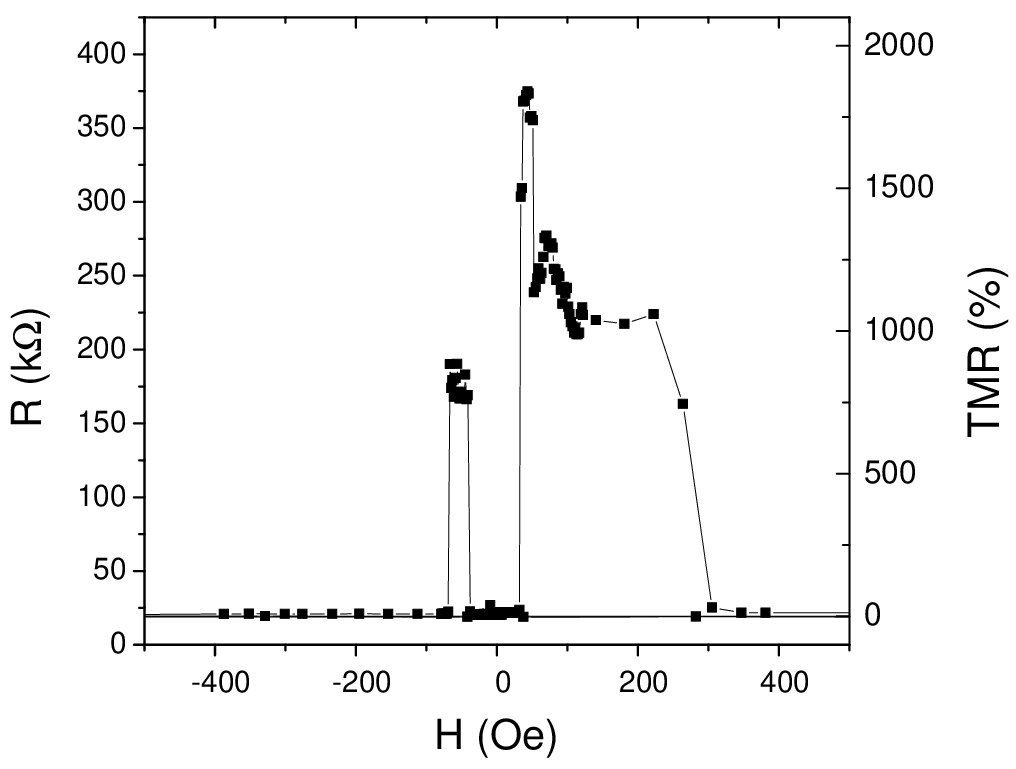}
 \caption{R(H) loop for a 5.6 $\mu$m $\times$ 5.6 $\mu$m
 junction measured at 4.2K.}
 \label{1800}
\end{figure}

Transport measurements were carried out in a four-point
measurement configuration. The resistance of the bottom LSMO
electrode was always at least one order of magnitude smaller than
that of the junction so that an artificial TMR enhancement due to
non-homogeneous current injection may be discounted. The LSMO
bottom electrode resistivity was in the 100-120 $\mu \Omega$ cm
range at T=4K , i.e. only about twice that of high-quality
epitaxial films and single crystals and a seven-fold reduction
compared to our previous process. We attribute this improvement
in LSMO electrode quality to switching to a neutralized ion beam
to etch the oxide layers and to cooling the sample during the
etching process \cite{sun2001}. Given that LSMO thin films are
prone to desorbing oxygen with rising temperature \cite{dorr2000},
limiting all sources of sample heating should reduce structural
(and therefore magnetic and electronic) modifications of the LSMO
electrodes and any possible interfacial oxygen diffusion between
LSMO and STO.

In Figure \ref{1800} we show the magnetic field dependence of the
resistance for a 5.6 $\mu$m $\times$ 5.6 $\mu$m junction, measured
at 4.2K after field-cooling and with a DC bias voltage
V$\rm{_{DC}}$=1 mV. We recall that a Co/CoO coverage overlayer
has been introduced to pin the top LSMO layer, so that, after
field-cooling, a symmetric variation is not expected. When
sweeping the field from negative to positive values, the
resistance of the junction rises from 19 k$\Omega$ to 375
k$\Omega$, yielding a TMR ratio of \mbox{1850 \%} with a field
sensitivity approaching 700 \%/Oe. From equation \ref{julliere}
and taking P=P$_1$=P$_2$, this value leads to a spin polarization
of $\rm{P}\simeq$ 95 \%. Four of our junctions showed a TMR larger
than 800 \% at 4K and 1 mV, i.e. P$>$89 \%. In addition, when
increasing applied bias, the TMR decreases more rapidly than in
standard MTJs with transition metal electrodes, which may be due
to stronger magnon scattering \cite{guinea98,bowenlsmotmrv2002}.
This suggests that a higher TMR ratio could be obtained if it
were possible to measure the junction at lower applied bias.

The asymmetry in the field dependence of the TMR for this
junction can be due to the shift of the magnetization reversal of
the pinned layer. Upon increasing the magnetic field this
separates the reversal fields of the two electrodes and leads to a
well-defined and high TMR peak. In decreasing field, the two
reversal fields are in the same range and the TMR is reduced.
This emphasizes the need for a uniaxial anisotropy to stabilize
well-defined antiparallel states like in the
La$_{2/3}$Ca$_{1/3}$MnO$_3$ / NdGaO$_3$ /
La$_{2/3}$Ca$_{1/3}$MnO$_3$ system studied by Jo \emph{et al}
\cite{jo2000}.

\begin{figure}
 \includegraphics[width=0.9\columnwidth]{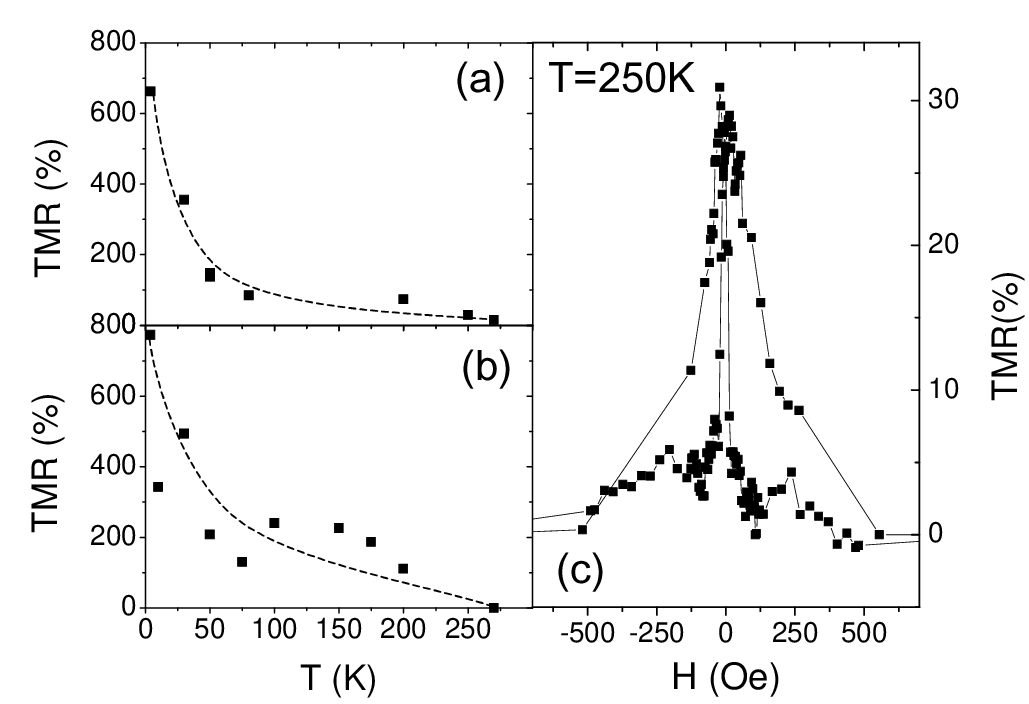}
 \caption{Temperature dependence of the TMR measured with
V$\rm{_{DC}}$=10 mV  for two junctions : 2 $\times$ 6 $\mu$m$^2$
(a) and a 1.4 $\times$ 4.2 $\mu$m$^2$ (b) (dashed lines are
guides to the eye). R(H) loop at T=250K and V$\rm{_{DC}}$=10mV
showing 30\% TMR (c).}
 \label{tmr_t}
\end{figure}

The temperature dependence of the TMR is plotted in Figure
\ref{tmr_t}(a, b) for two junctions using R(H) loops measured at
V$\rm{_{DC}}$=10 mV. The TMR decreases rather quickly upon
increasing T but only vanishes at temperatures of some 280K for
the 2x6 $\mu$m$^2$ junction. TMR ratios of 30 \% (Figure
\ref{tmr_t}(c)) and 12 \% are obtained at 250K and 270K,
respectively. This represents a sizeable improvement with respect
to previous results \cite{viret97} obtained from heterostructures
grown in identical conditions. Since this system shows a
temperature dependent competition between the junction dipolar
field, the shape anisotropy and the CoO pinning effects described
previously, it is difficult to stabilize a fully antiparallel
alignment when temperature increases. This leads to a sharp
extrinsic decrease of the TMR with temperature. Finally, since
these results were obtained in LSMO/STO/LSMO junctions with a
fully strained crystallographic structure, we state that strain
is not a limiting factor toward conceiving manganite-based MTJs
with sizeable TMR values at relatively high temperatures,
contrary to what was suggested by Jo \emph{et al} \cite{jo2000}.
High quality interfaces which limit the disruption of the
manganite's properties fulfill a more important requisite, as
suggested by previous studies \cite{bibes2002}.

In summary, we have observed a magnetoresistance of 1850 \% in
LSMO-based tunnel junctions, from which we deduce an average spin
polarization of at least 95 \% for LSMO at the interface with
STO. This value may be higher at lower temperature and junction
bias. As such, this result - the highest spin polarization
measured in any material from tunneling experiments - underscores
the transport half-metallic nature of mixed-valence manganites. In
addition, the temperature dependence of the magnetoresistance for
these junctions is better than previously reported, as the TMR
vanishes only at about 280K. We attribute these improved results
mainly to the use of an upgraded lithographic process which
defines micron-sized tunnel junctions and strongly limits sample
heating. Our findings show that strained LSMO can indeed be used
as a spin analyzer to perform fundamental tunneling studies,
possibly up to room temperature, as well as a source of fully spin
polarized current in epitaxial oxide heterostructures.

\acknowledgments{}
 We are deeply indebted to Eric Jacquet and Annie Vaures for sample growth, and to Josette Humbert for some sample
 processing. We acknowledge financial support within the European Union AMORE project.

\end{document}